\documentclass[runningheads]{llncs}

\usepackage{graphicx}
\usepackage{amsmath}
\usepackage{amssymb}
\usepackage{booktabs}
\usepackage{xcolor}
\usepackage{multirow}
\usepackage{todonotes}
\usepackage{tabularx}
\usepackage{url}
\usepackage{comment}
\usepackage{adjustbox}

\newcommand{\printfnsymbol}[1][\value{footnote}]{\footnotemark[#1]}
\usepackage{authblk}
\begin{document}
\title{Spatio-temporal Learning from Longitudinal Data for Multiple Sclerosis Lesion Segmentation}
\titlerunning{Spatio-temporal Learning for Longitudinal Segmentation}

\author{Stefan Denner\inst{1,}\thanks{\textit{First two authors contributed equally to this work.
\newline $^\dagger$ S. T. Kim and N. Navab share senior authorship.
\newline $^{\ast\ast}$ Corresponding author (seongtae.kim@tum.de)}}, 
Ashkan Khakzar\inst{1,}\printfnsymbol, 
Moiz Sajid\inst{1}, 
Mahdi Saleh\inst{1}, 
Ziga Spiclin\inst{3}, 
Seong Tae Kim\inst{1,\dagger,\ast\ast}, 
Nassir Navab\inst{1,2,\dagger}}
\authorrunning{Denner and Khakzar et al.}
\institute{Computer Aided Medical Procedures, Technical University of Munich, Germany
\and
Computer Aided Medical Procedures, Johns Hopkins University, USA
\and
Faculty of Electrical Engineering, University of Ljubljana, Ljubljana, Slovenia}

\maketitle              
\begin{abstract}
Segmentation of Multiple Sclerosis (MS) lesions in longitudinal brain MR scans is performed for monitoring the progression of MS lesions.
We hypothesize that the spatio-temporal cues in longitudinal data can aid the segmentation algorithm. Therefore, we propose a multi-task learning approach by defining an auxiliary self-supervised task of deformable registration between two time-points to guide the neural network toward learning from spatio-temporal changes.
We show the efficacy of our method on a clinical dataset comprised of 70 patients with one follow-up study for each patient. Our results show that spatio-temporal information in longitudinal data is a beneficial cue for improving segmentation. We improve the result of current state-of-the-art by 2.6\% in terms of overall score (\textit{p}$<$0.05). 
Code is publicly available\footnote{{\color{purple}\url{https://github.com/StefanDenn3r/Spatio-temporal-MS-Lesion-Segmentation}}}.

\end{abstract}
\section{Introduction}

%
%
Multiple Sclerosis (MS) is a neurological disease characterized by damage to myelinated nerve sheaths (demyelination) and is a potentially disabling disease of the central nervous system. The affected regions appear as focal lesions in the white matter~\cite{Steinman1996} and
Magnetic Resonance Imaging (MRI) is used to visualize and detect the lesions~\cite{Compston2008}. MS is a chronic disease, 
therefore longitudinal MRI patient studies are conducted to monitor the progression of the disease.
Accurate lesion segmentation in the MRI scans is important to quantitatively assess response to treatment~\cite{stangel_towards_2015} and future disease-related disability progression \cite{uher_combining_2017}. However, manual segmentation of MS lesions in MRI volumes is time-consuming, prone to errors and intra/inter-observer variability~\cite{carass2017longitudinal}.

Several studies have proposed automatic methods for MS lesion segmentation in MRI scans~\cite{ghafoorian2015convolutional,andermatt2017automated,valverde2017improving,hashemi2018asymmetric,zhang2019multiple,aslani2019multi}.
Valverde et al.~\cite{valverde2017improving} proposed a cascade of two 3D patch-wise convolutional networks, where the first network provides candidate voxels to the second network for final lesion prediction. 
Hashemi et al.~\cite{hashemi2018asymmetric} introduced an asymmetric loss, similar to the Tversky index, which is supposed to tackle the problem of high class imbalance in MS lesion segmentation by achieving a better trade-off between precision and recall. 
Whereas the previous two approaches worked with 3D input, Aslani et al.~\cite{aslani2019multi} proposed a 2.5D slice-based multimodality approach, where they use a single branch for each modality. They trained their network with slices from all plane orientations (axial, coronal, sagittal). During inference, they merge those 2D binary predictions to a single lesion segmentation volume by applying a majority vote. 
Zhang et al.~\cite{zhang2019multiple} also proposed a 2.5D slice-based approach, but they concatenated all modalities instead of processing them in multiple branches. In contrast, they utilize a separate model for each plane orientation. 


However, none of these works use the data from multiple time-points.
The work of Birenbaum et al.~\cite{birenbaum2016longitudinal} is the only method that processes longitudinal data. Birenbaum et al.~\cite{birenbaum2016longitudinal} proposed a siamese architecture, where input patches from two time-points are given to separate encoders that share weights and subsequently the encoders' outputs are concatenated and fed into subsequent CNN to predict the class of pixel of interest. Birenbaum et al.~\cite{birenbaum2016longitudinal} set the direction for using longitudinal data and opens up a line of opportunities for future work.
However, their work does not extensively investigate the potential of using information from longitudinal data. 
Specifically, their proposed late-fusion of features does not properly take advantage of learning from structural changes.

In this paper, we initially propose an improved baseline methodology over that of Birenbaum et al.~\cite{birenbaum2016longitudinal} by employing an early fusion of multimodal longitudinal data, which allows for proper capturing of the differences between inputs from different time points, as opposed to the late fusion of data proposed in~\cite{birenbaum2016longitudinal}.
Our main contribution is \emph{proposing a multitask learning framework by adding an auxiliary deformable registration task to the segmentation model}. The two tasks share the same encoder, but two separate decoders are assigned for each task, hence the learned features for predicting the deformation are shared with the segmentation task. The notion of joint registration and segmentation itself is previously proposed in \cite{xu2019deepatlas} (Deep Atlas), however, the methodology uses two different segmentation and registration networks where there is no feature sharing in between the models, and the output of segmentation model for two different and single-time point scans are corrected by the deformation and vice versa. On the contrary, our proposed approach aims to use the learned features for the registration task \emph{explicitly} in the segmentation task, and to our knowledge this is the first work which applies such a model to longitudinal data.
We hypothesize that structural changes of lesions through time are valuable cues for the model to detect these lesions, and we evaluate our approaches on a clinical dataset including 70 patients with 
one follow-up study for each patient. 
We compare our methods to the state-of-the-art works on MS lesion segmentation~\cite{zhang2019multiple,hashemi2018asymmetric}, and the previous longitudinal approach~\cite{birenbaum2016longitudinal}. Moreover, we adapt the joint registration and segmentation methodology of~\cite{xu2019deepatlas} to longitudinal data and the problem of MS lesion segmentation, and compare it to our own methodology to investigate how explicitly incorporating spatio-temporal features can improve the segmentation. 




\section{Methodology}
\begin{figure}[t]
\centering
  \includegraphics[width=0.8\textwidth]{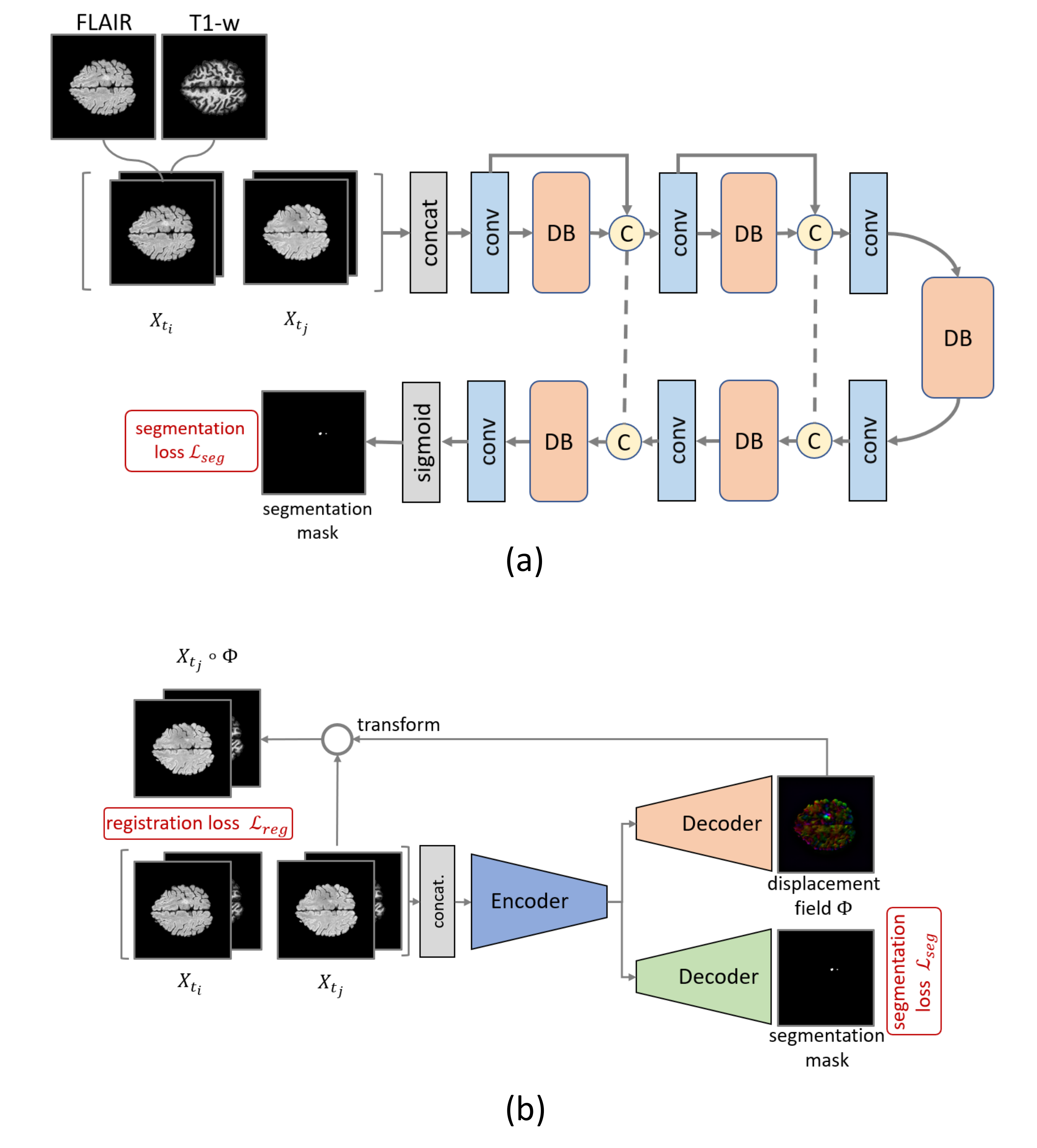}
  \caption{Our proposed methods: (a) Baseline Longitudinal Network: longitudinal scans are concatenated and given to the segmentation model to implicitly use the structural differences (b) Multitask Longitudinal Network: The network is trained with an auxiliary task of deformable registration between two longitudinal scans, to explicitly guide the network toward using spatio-temporal changes.}
  \label{fig:methods}
\end{figure}
This section describes our approaches for incorporating spatio-temporal features into the learning pipeline of a neural network.
We hypothesize that structural changes of lesions between the longitudinal scans are valuable cues for detecting these lesions. Note that the aim is not to model \emph{how} the lesions deform or change, but to find \emph{what} has changed and to use that information to improve segmentation. To this aim, we propose a baseline neural network that improves on the methodology of~\cite{birenbaum2016longitudinal} by proposing early-fusion of input data and subsequently introduce our multitask learning approach with deformable registration approach.
%
%
%
%
%
%
\subsection{Baseline Longitudinal Network}\label{method:long_arch}
%
It is shown that state-of-the-art results can be achieved using the 2.5D approach~\cite{zhang2019multiple}. For the case of the 3D approach, it is challenging to directly process a full 3D volume by the current available GPU memory \cite{roy2019quicknat}. Therefore, 3D models are usually operated on patches extracted from the volume \cite{wachinger2018deepnat,hashemi2018asymmetric}, which limits the context for accurate prediction.  
Thus, we adopt the 2.5D approach~\cite{Roth2014,aslani2019multi,zhang2019multiple} for segmentation of 3D MR volumes (we report the inference time required for the segmentation of each scan using 2.5D approach in section~\ref{sec:implementaiton}.). For each voxel, segmentation is done on the three orthogonal slices crossing the voxel of interest. The probability output of the corresponding pixel in each view is averaged and thresholded to determine the final prediction for the voxel.
To segment a given slice, we use a fully convolutional and densely connected neural network (Tiramisu)~\cite{jegou2017one}. The network receives a slice from any of the three orthogonal views and outputs a segmentation mask. To account for different modalities (T1-w, FLAIR), we stack the corresponding slices from all modalities and feed them to the network. 

In order to use the structural changes between the two time-points, we give the concatenated scans of the two-time points as input to the segmentation network (Fig.~\ref{fig:methods}.a).
This early-fusion of inputs allows the network filters to capture the minute structural changes at all layers leading to the bottleneck, as opposed to the late fusion of Birenbaum et al.~\cite{birenbaum2016longitudinal}, where high-level representations from each time point are concatenated. The early fusion's effectiveness for learning structural differences can be further supported by the similar architectural approaches in the design of deformable registration networks~\cite{Balakrishnan2018,Balakrishnan2019}.

%
\subsection{Multitask Learning with Deformable Registration}\label{method:multitask}

In this section we describe our approach involving the augmentation of the segmentation task with an auxiliary deformable registration task. We aim to explicitly use the structural change information between the two longitudinal scans. In longitudinal scans, only specific structures such as MS lesions change substantially.
Deformable registration is defined to learn a deformation field between two instances. We therefore propose augmenting our baseline longitudinal segmentation model with a deformable registration loss.
We hypothesize that this would further guide the network towards using structural differences between the inputs of two different time points. Note that the longitudinal scans are already rigidly registered in the pre-processing step, therefore the deformation field only reflects the structural differences of lesions. 

The resulting network (Fig.~\ref{fig:methods}.b) consists of a shared encoder followed by two decoders used to generate the specific outputs for the two tasks. One head of the network is associated with generating the segmentation mask and the other one with deformation field map. The encoder-decoder architecture here is that of Tiramisu~\cite{jegou2017one}, and two decoders for registration and segmentation are architecturally equivalent. The deformable registration task is trained without supervision or ground truth registration data. This is rather trained self-supervised and by reconstructing one scan from the other which helps adding additional generic information to the network. The multi task loss is defined as:
\begin{equation}
\mathcal{L} = \mathcal{L}_{seg} + \mathcal{L}_{reg}
\end{equation}
A common pitfall in multi task learning is the imbalance of different tasks which leads to under-performance of multitask learning compared to single tasks. To solve this one needs to normalize loss functions or gradients flow~\cite{Chen2018}. Here we use the same type of loss function for both tasks. Specifically we use MSE loss, which is used for both registration and segmentation problems. We use a CNN based deformable registration methodology similar to VoxelMorph~\cite{Balakrishnan2019}, but adapted to 2D inputs and using Tiramisu architecture (Fig.~\ref{fig:methods}.b). The registration loss ($\mathcal{L}_{reg}$) is defined as:
\begin{equation}
\mathcal{L}_{reg} = \mathcal{L}_{sim}(X_{t_{i}},X_{t_{j}}\circ \Phi) + \lambda \mathcal{L}_{smooth}(\Phi)
\end{equation}
where $\Phi$ is the deformation field between inputs $X_{t_{i}}$ and $X_{t_{j}}$ ($t_{i}$ and $t_{j}$ denote the time-points). $X_{t_{j}}\circ \Phi$ is the warping of $X_{t_{j}}$ by $\Phi$, and $\mathcal{L}_{sim}$ is the loss imposing the similarity between $X_{t_{i}}$ and warped version of $X_{t_{j}}$. $\mathcal{L}_{smooth}$ is regularization term to encourage $\Phi$ to be smooth. We use MSE loss for $\mathcal{L}_{sim}$, and for the smoothness term $\mathcal{L}_{smooth}$, similar to~\cite{Balakrishnan2019} we use a diffusion regularizer on the spatial gradients of the displacement field.

\section{Experiment Setup}
\subsection{Datasets and Preprocessing}
The clinical dataset \cite{galimzianova2016stratified,lesjak2018novel} consists of 1.5T and 3T MR images. Follow-up images of 70 MS patients were acquired. Images are 3D T1-w MRI scans and 3D FLAIR scans with approximately 1 mm isotropic resolution.
The MR scans were preprocessed initially by applying non-local mean-based image denoising~\cite{manjon201218} and the N4 bias correction~\cite{tustison2010}.
From the preprocessed T1-weighted (T1) image the brain mask was extracted by a multi-atlas label fusion segmentation method~\cite{cardoso2015}, which employed 50 manually segmented T1 MR brain images of age-matched healthy subjects.
The atlases were aligned to the brain masked and preprocessed T1 by a nonlinear B-spline registration method~\cite{klein2010}. 
Using the same registration method, the corresponding preprocessed FLAIR images were aligned to the preprocessed T1.
MS lesions were manually annotated by an expert rater. The data of 40 patients were used as a training set (30 patients for training the model and 10 patients for validation).
The remaining data of 30 patients were used as an independent test set.


\subsection{Implementation Details}\label{sec:implementaiton}
The encoders and decoders of our architectures are based on FC-DenseNet57~\cite{jegou2017one}.
We used Adam optimizer with AMSGrad~\cite{reddi2019convergence} and a learning rate of 1e-4. 
We use a single model with shared weights for all plane orientations. Since our approaches are 2.5D, we average and threshold the probability output predictions of all plane orientations. The inference time for a whole 3D volume with a resolution of 224$\times$224$\times$224 is 8.29 seconds on a 8 GB GPU. Our Multitask  Longitudinal Network has about 2 million parameters. PyTorch 1.4~\cite{paszke2019pytorch} is used for neural network implementation.

\subsection{Evaluation Metrics}
For evaluating our methods, we use Dice Similarity Coefficient (DSC), Positive Predictive Value (PPV), Lesion-wise True Positive Rate (LTPR), Lesion-wise False Positive Rate (LFPR) and Volume Difference (VD). To consider the overall effect of different metrics, we adopt an \emph{Overall Score} in a similar way to MS lesion segmentation challenges~\cite{styner20083d,carass2017longitudinal} as follows:

\begin{equation} \label{Overall_score}
\begin{split}
    Overall Score 
    &= 0.125DSC+0.125PPV+0.25(1-VD) \\
    &+0.25LTPR+0.25 (1-LFPR).
\end{split}
\end{equation}

\subsection{Method Comparisons}\label{sec:method_comparisons}
In this section we elaborate upon the methods used for comparison and the naming scheme used throughout the paper. Henceforth we call the single time point segmentation models as \emph{static} models.
All models are based on the FC-DenseNet-57 architecture to provide comparability.

\noindent \textbf{Static Network (Zhang et al. \cite{zhang2019multiple})}: 
The approach from Zhang et al. \cite{zhang2019multiple} feeds three consecutive slices of one 3D scan to the network. It uses one model for each plane orientation, i.e. weights are not shared along the different plane orientations. 

\noindent \textbf{Static Network (Asymmetric Dice Loss \cite{hashemi2018asymmetric})}:
Hashemi et al. \cite{hashemi2018asymmetric} proposed an asymmetric dice loss which is applied on the 
FC-DenseNet-57. $\beta$ = 1.5

\noindent \textbf{Deep Atlas \cite{xu2019deepatlas}}: Two separate FC-DenseNet-57 which are jointly trained. One model is trained on the deformation task, the other on is trained on segmentation. 


\noindent \textbf{Longitudinal Siamese Network ~\cite{birenbaum2016longitudinal}}: We implement the longitudinal siamese model~\cite{birenbaum2016longitudinal} with a FC-DenseNet-57 \cite{jegou2017one}. As in \cite{birenbaum2016longitudinal}, we use the late fusion which combines the features of different time point scans in the bottleneck layer.

\noindent \textbf{Baseline Static Network (ours)}: Similar to Zhang et al.~\cite{zhang2019multiple}, our model is based on FC-DenseNet-57 and uses only a \emph{single} time-point. Our model is trained with all three plane orientations (axial, coronal, and sagittal). 

\noindent \textbf{Baseline Longitudinal Network (ours)}: Our proposed longitudinal model (section~\ref{method:long_arch}).

\noindent \textbf{Multitask Longitudinal Network (ours)}: Our multitask network (section~\ref{method:multitask}).


\section{Results and Discussion}
\begin{figure*}[t]
	\centering
	\includegraphics[width=0.98\textwidth]{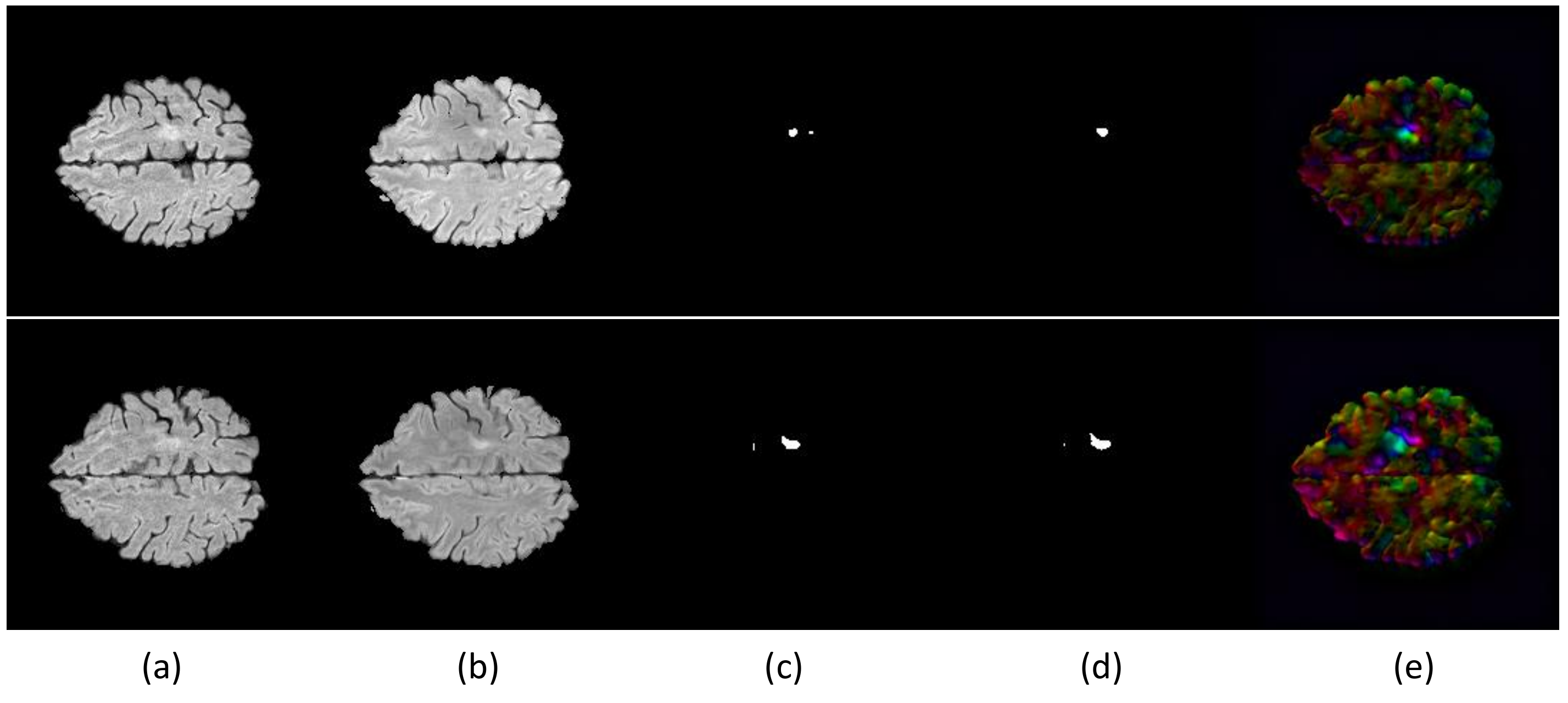}
	\caption{Visualisation of MS lesion's structural change in two longitudinal MR FLAIR scans. Each row presents data from one patient. (a) is the scan from the first time-point, (b) is the scan from the follow up study, (c) visualizes ground truth of MS lesions on the follow up image, (d) shows the predicted segmentation mask of Multitask Longitudinal Network on the follow up image, (e) represents the predicted displacement field between the two scans using the registration module of our multitask method.}
	\label{Figure_Qualitative} 
\end{figure*} 
\begin{table}[t]
	\centering
	\caption{Comparison of different methodologies. Our methods are shown in bold letters. For LFPR and VD, lower is better. Standard deviations are provided in the supplementary materials.}
    {

    \resizebox{\linewidth}{!}{\setlength{\tabcolsep}{4pt}
    \begin{tabular}{c|c|c|c|c|c|c}
        \toprule
        \multirow{2}{*}{\textbf{Method}} & \multirow{2}{*}{\textbf{DSC} $\uparrow$} &
        \multirow{2}{*}{\textbf{PPV} $\uparrow$} &  \multirow{2}{*}{\textbf{LTPR} $\uparrow$} & \multirow{2}{*}{\textbf{LFPR} $\downarrow$} & \multirow{2}{*}{\textbf{VD} $\downarrow$}
        & \multirow{2}{*}{\textbf{Overall Score} $\uparrow$}\\  
        & &  & 
        & &  & \\ \midrule  
        \textbf{Multitask Longitudinal Network} & $\textbf{0.695}$ & $\underline{0.771}$ & $\underline{0.680}$ & $\underline{0.212}$ &  $\underline{0.221}$ & $\textbf{0.745}$\\ 
        
        \textbf{Baseline Longitudinal Network} & $\underline{0.694}$  & 0.752 & 0.654 & 0.227 & 0.227 & $\underline{0.731}$ \\ 
        
        \textbf{Baseline Static Network} &  0.684 & 0.762 & 0.647 & 0.250 & 0.247 & 0.718\\  
        Longitudinal Siamese Network \cite{birenbaum2016longitudinal}  & 0.684  & $\textbf{0.777}$ & 0.614 & $\textbf{0.194}$ & 0.245 & 0.726  \\  

        Static Network (Zhang et al. \cite{zhang2019multiple}) &  0.684   &  0.761 & 0.604 & 0.223 & 0.263 & 0.710\\  
        
        Static Network (Asymmetric Dice Loss \cite{hashemi2018asymmetric}) &  0.690   &  0.648 & $\textbf{0.752}$ & 0.346  & 0.336  & 0.685 \\  
        
        Deep Atlas \cite{xu2019deepatlas} &  0.656   &  0.701 & 0.652 & 0.260  & $\textbf{0.180}$  & 0.723 \\  
        \bottomrule
    \end{tabular}}} \label{Table_Comparison_Main}
\end{table}
We first compare our Baseline Static Network (Sec~\ref{sec:method_comparisons}) to the state-of-the-art static segmentation approaches~\cite{zhang2019multiple,hashemi2018asymmetric}. Table~\ref{Table_Comparison_Main} shows that our Baseline Static Network, which is inspired by the method of~\cite{zhang2019multiple} achieves an overall score of 0.718  and performs similar to our implementation of~\cite{zhang2019multiple} which achieves 0.710. The difference between the methods is that we use one single network for all orthogonal views, and the results show that using a single network and thus, fewer parameters achieves similar but slightly better results. We also report the results of asymmetric Dice loss of~\cite{hashemi2018asymmetric} on our static model which improves the DSC, however hurts the Overall Score.

We proceed by evaluating our Baseline Longitudinal Network against our Baseline Static Network and the Longitudinal Siamese Network of ~\cite{birenbaum2016longitudinal}. In Table~\ref{Table_Comparison_Main} we observe that both longitudinal models improve the overall score. We also observe that our Baseline Longitudinal Network which uses early fusion of input data improves the score further. In section~\ref{method:long_arch}, we stated that the early-fusion of inputs allows the network to capture the structural differences between the inputs better than late-fusion. The comparison between our Longitudinal Network and Longitudinal Siamese Network~\cite{birenbaum2016longitudinal} which only differ in how they fuse the inputs, serves as an ablation experiment for this claim.

To illustrate the behavior of our Multitask Longitudinal Network, we visualize the segmentation mask and the displacement field in Fig. \ref{Figure_Qualitative}. The displacement field shows \emph{what} has changed. In Fig. \ref{Figure_Qualitative}, the colors in the displacement encode the direction of the field at any point, and the brightness signifies the magnitude of displacement. As can be seen, the areas corresponding to MS lesions have high brightness indicating that the deformable registration model has captured the change of MS lesions.
As shown in Table~\ref{Table_Comparison_Main}, by exploiting the spatio-temporal information from deformable registration, the segmentation performances were improved compared to the Baseline Longitudinal Network. It is also observed that the Deep Atlas~\cite{xu2019deepatlas} methodology while improving over the static approaches, achieves inferior results compared to all longitudinal approaches.

\begin{table}[t]
	\centering
	\caption{Statistical significance analysis of performance improvements of our Multitask Longitudinal Network over other methods in terms of the overall score by paired t-test.}
    {
    \resizebox{.9\linewidth}{!}{\begin{tabular}{c|c|c|c}
        \toprule
        \multirow{2}{*}{\textbf{Comparison with}} & \textbf{Mean difference}  
        & \multirow{2}{*}{\textbf{95\% CI}} & \multirow{2}{*}{\textbf{\textit{p}-value}} \\
         & \textbf{$\pm$ standard error} &  & \\ \midrule
        Baseline Longitudinal Network & 0.0141$\pm$0.0047    &  [0.0047,0.0236]  & 0.0040   \\  
        
        Baseline Static Network &   0.0267$\pm$0.0053    &  [0.0161,0.0372]  & $<$0.0001 \\  
        
        Longitudinal Siamese Network  \cite{birenbaum2016longitudinal}  &   0.0187$\pm$0.0048    &  [0.0091,0.0282]  & 0.0002 \\  
        
        Static Network (Zhang et al. \cite{zhang2019multiple}) &   0.0389$\pm$0.0055    &  [0.0280,0.0499]  & $<$0.0001 \\  
        
        Static Network (Asymmetric Dice Loss \cite{hashemi2018asymmetric})  &   0.0603$\pm$0.0144    &  [0.0315,0.0891]  & 0.0001 \\  
         
        Deep Atlas \cite{xu2019deepatlas} &   0.0223$\pm$0.0094    &  [0.0034,0.0412]  & 0.0218 \\  
        
        \bottomrule
    \end{tabular}}} \label{Table_StatisticalAnalysis}
\end{table}

To verify that the performance improvements of our Multitask Longitudinal Network are statistically signiﬁcant, we conducted further analysis of the models with paired \textit{t}-test on the overall score. The paired \textit{t}-test provides a statistical evaluation of the performance differences between models. Table~\ref{Table_StatisticalAnalysis} shows the results of the statistical significance analysis for differences of overall score. 
The multitask learning framework significantly improves the overall score compared with longitudinal network (\textit{p}=0.004). The improvements from our Multitask Longitudinal Network compared with previous methods \cite{birenbaum2016longitudinal,zhang2019multiple,xu2019deepatlas,hashemi2018asymmetric} are also statistically significant $p<0.05$. 

\section{Conclusion}
In this work, we investigated the utilization of spatio-temporal information in longitudinal brain MR data to improve the segmentation of MS lesions. We proposed a novel multitask formulation where an auxiliary unsupervised deformable registration task is adopted. We evaluated our approaches on a clinical dataset comprising of 70 patients with one follow-up scan for each patient. Our evaluations against state-of-the-art MS lesion segmentation works confirm that incorporating spatio-temporal information into segmentation models improves the segmentation performance. Furthermore, we showed transferring previous work on joint registration and segmentation to longitudinal data achieves inferior results compared to our methodology as we explicitly incorporate spatio-temporal features into our model. In future work, our proposed methodology can be extended to other longitudinal medical studies to improve segmentation.

\section*{Acknowledgements}
\label{sec:acknowledgements}
The authors acknowledge the financial support for this work by Siemens Healthineers and Munich Center for Machine Learning (MCML). Ziga Spiclin was supported by the Slovenian Research Agency (research core funding No. P2-0232, and research grant No. J2-2500).


\bibliographystyle{splncs04}
\bibliography{main}



\end{document}